# Statistical Thermodynamics of Strongly Coupled Plasma

A.H. Khalfaoui and D. Bennaceur-Doumaz

*CDTA/ Division des Milieux Ionisés et Laser, Houch Oukil BP 17, Baba Hassen, Algiers/Algeria*

**Abstract**: To analyze nonidealities inherent to degenerate plasma, a quantum collective approach is developed. Thermodynamic functions of a system of partially degenerate electrons and strongly coupled ions are derived from first principles. The model takes into account the energy eigenvalues of i) the thermal translational particle motions, ii) the random collective electron and ion motions, and iii) the static Coulomb interaction energy of the electrons and ions in their oscillatory equilibrium positions. These statistical thermodynamic calculations lead to simple analytical expressions for internal energy as well as an equation of state (EOS). A dispersion relation for the high frequency branch of the plasma oscillations is introduced to take into account the partial degeneracy character and thereby to quantify temperature finiteness effects on thermodynamic properties of a partially degenerate plasma. The present results are in good quantitative agreement with the existing models. These latter being based mainly on numerical experiments while in the present model more physical insight is explicitly stated. This makes a contribution to the theoretical knowledge of coupled plasma for thermonuclear fusion as well as of astrophysical interests.

## I. INTRODUCTION

Thermodynamic properties of high density plasma with completely or partially degenerate electrons and strongly coupled ions are a subject of great interest, where more refinements are still needed. Strongly coupled plasmas are currently produced by heavy ion beam interaction with solid targets, with densities close to those of solid state [1]. Recent equation of state (EOS) measurements, with a variety of techniques, have been obtained for liquid deuterium and thoroughly discussed [2-4].

The understanding and the interpretation of these experimental results require a fundamental and rigorous approach which must take into account the main character of the medium such as strong coupling, degeneracy as well as quantum effects.

Several models have been developed with different assumptions and applied to determine thermodynamic as well as the static and dynamic correlation functions, where the physical situation has, often, been described by the simple model of the classical one-component plasma (OCP) [5,6]. Even though the OCP model may explain many important characteristics of dense ionized matter, but for some domain of physical parameters one has to take into account some nonidealities, as for instance the polarization of electrons, which induces the screening of ionic charges while static and dynamic properties of the system are thereby modified from those of the OCP.

Various investigations have been done to analyze the nonideal or high density plasmas properties where the Monte Carlo numerical experiments were the mostly performed approaches [6-9], to which some theoretical models are being developed and compared [10].

Moreover, the density fluctuations for dense plasmas can be split into two approximately independent components associated, respectively, with the collective and individual aspects of the system. The collective component, which is present only for wavelengths greater than Debye length, represents organized oscillations, brought about by the long-range part of the Coulomb interactions [11]. When such an oscillation is excited, each individual particle suffers a small perturbation of its velocity and position, arising from the combined potential of all other particles. The contribution to the density fluctuations resulting from these perturbations is in phase with the potential producing it, so that in an oscillation it is found an organized wave-like perturbation superposed on the random thermal motion of the particle.

For the high density, partially or completely degenerate plasma, i.e., at relatively low temperature, the thermal motion no longer plays the dominant role. Instead the cumulative potential of all particles will be considerable because the range of the force permits to a very large number of particles to contribute to the potential at a given point. Hence the collective aspect would be dominant and particularly governs the thermodynamic properties of the plasma in consideration here.

In addition this aspect of collective approach has proven to be a very efficient model to govern the transport phenomena and the relaxation processes in strongly coupled plasma [12]. For these reasons, a quantum statistical theory will be presented here, for strongly coupled plasma, based on concepts similar to those used by Debye for solids [13] and carry this approach further to obtain a quantification of thermodynamic quantities, while including several nonideal effects.

The role of the longitudinal phonons of the Debye theory is played here by the quanta of the plasma oscillations (plasmons and ion sound waves).

Moreover, even though zero temperature calculations may be useful for many applications especially for evaluating the chemical composition of condensed matter, they cannot be applied to quantify some parameters at finite temperature, since they do not provide information about thermal properties such as heat capacity or adiabatic exponents. Furthermore, a characteristic and important feature of high density plasma is precisely the involvement of a wide range of Fermi degeneracy, i.e., with finite electron temperature.

For that, dispersion relations as functions of both, the temperature and the Fermi energy, i.e., the density, will be introduced in the present work, and will make possible the calculations of the thermodynamic functions of partially degenerate plasma.

Moreover, in order to complete the whole picture of the present model, it should be noted as stated above, that the collective motion of the particles is superposed on their individual random thermal motion. This approach to be applied to dense plasmas is justified since the medium can be described as distorted lattice. The ions are kept in positions with lattice-like structure due to the strong interparticle correlations, but with an incomplete ordering. The system being a fluid, the average position of the particles changes slowly with time unlike the real lattice that appears with stronger correlations [14].

## II. STATISTICAL THERMODYNAMICS

In the plasma under consideration, the electrons and ions interact through their longitudinal Coulomb fields (transverse electromagnetic interactions are negligible in the present model).
The resulting Hamilton function with Coulomb interactions gives the free energy of the system as:

$$F = F_0 + \Delta F, \qquad (1)$$

$$F_0 = \sum_{s=e,i} F_s^{(0)}; \quad \Delta F = \sum_{s=e,i} \widetilde{F}_s + E_M,$$

where $F_s^{(0)}$ is the ideal free energy of the non interacting plasma components, $E_M$ is the Coulomb interaction energy of the electrons and ions in their equilibrium positions, and $\widetilde{F}_s$ is the free energy of the electron and ion oscillations.

In the Hamilton function [Eq.(1)] the most significant, short and long range Coulomb interactions are taken into consideration for all types and species, e-i , e-e and i-i by means of the Madelung energy $E_M$ and the plasmon energy $\widetilde{F}_s$ for which $s = e$ corresponds to the high frequency branch and $s = i$ are the ion sound waves (low frequency branch).

Moreover, in equation (1) the main interest is to be given to $\Delta F$ since $F_0$, for partially or completely degenerate electrons are given by Tolman [15], where the ions behave in general classically.

The equilibrium positions of the electrons and ions, about which the electrostatic oscillations occur form a lattice-like with an incomplete ordering for each electron and each ion.

The Coulomb interaction energy being equivalent for both electron or ion lattice, the Madelung

energy is given by:

$$E_M = -N\kappa_B T \alpha(\Gamma)\Gamma, \qquad (2)$$

where $\alpha$ is a parameter depending on $\Gamma$ ($\alpha = \alpha(\Gamma)$). However it has been shown that for $r_s < 1$ the ordering of the plasma increases with $\Gamma$, $\alpha(\Gamma)$ becomes a weak function of the coupling parameter such that asymptotically $\alpha = \bar{\alpha}$ for all $\Gamma \gg 1$ and $\bar{\alpha} = 0.9$ deduced from the ion sphere model where $\Gamma$ is the coupling parameter ($\Gamma = (Ze)^2 / a_i \kappa_B T$ with $a_i = (3Z/4\pi n)^{1/3}$).
On the other hand for $\Gamma \ll 1$, equation (2) indicates, that $E_M / N$ is of the order of the average i-i energy and for such a low coupling, from Debye-Huckel model it has been shown that $\alpha \sim \Gamma^{1/2}$ and given, by:

$$\alpha(\Gamma) = \frac{1}{\sqrt{3}}(1+Z)^{3/2} Z^{-3} \Gamma^{1/2}. \qquad (3)$$

As to the free energy of the plasma oscillations, it can be deduced from statistical thermodynamic approach.
Since the plasma volume V contains N electrons and N/Z ions, there exist N (high frequency branch) and N/Z (low frequency branch) characteristic frequencies of longitudinal oscillations. The energy of the plasma state with $n=1,2,3\ldots$ plasmons of frequency $\omega_j$ is $E\{j\} = \sum_n n\hbar\omega_j$.

Accordingly the partition function Q of the longitudinal plasma oscillations is:

$$Q = \prod_j \sum_n \exp(-n\hbar\omega_j / \kappa_B T) = \prod_j \{1 - \exp(\hbar\omega_j / \kappa_B T)\}^{-1}, \qquad (4)$$

and the free energy $\tilde{F}$ of the plasmons is:

$$\tilde{F} = -\kappa_B T \ln Q = \kappa_B T \sum_j \{1 - \exp(\hbar\omega_j / \kappa_B T)\}, \qquad (5)$$

and all the thermodynamic functions such as, the equation of state, the internal energy, etc… can now be deduced.
In the limit $V \to \infty$, the discrete eigenfrequencies $\omega_j$ will be replaced by a continuum spectrum $\omega_j = \omega_s(q), (s=e,i)$ in accordance with the dispersion law for space charge waves of wavelengths

$\lambda = 2\pi / q$, with $0 \leq q \leq \hat{q}$. And hence the problem now lies in the choice of the dispersion relations $\omega_s(q)$.

## III RESULTS AND COMPARISONS

In this section, the results obtained by the present method will be analyzed and also compared to existing approaches. First of all, the major results of the present work are analytical expressions for free energy from which all the thermodynamic functions can be derived. Internal energy as well as the equation of state are explicitly reported for dense two component plasma over a wide range of density and temperature.

Thermodynamic functions derived, through a relatively simple procedure, will be studied in the context of partial degeneracy through the variable $\theta$, the coupling strength $\Gamma$, ($\theta = \kappa_B T / E_F$ where $\kappa_B$ is the Boltzmann constant, T the temperature and $E_F$ the Fermi energy).

As it will be seen, the present results compare very well with the existing theories, with more physical insight allowed by the statistical thermodynamic approach of the present calculations.

The variety of effects described in the present collective approach are the dynamic screening, the incomplete degeneracy (temperature finiteness) through the development of a dispersion relation depending on both the Fermi energy as well as the temperature and the quantum effects through the general character of the model which considers the plasma as an ensemble of elementary excitations that are quasi-particles.

By analogy to most existing models, the Madelung contribution to the thermodynamic functions of the present theory, should behave as the OCP contribution while the high and low frequency branches oscillations play the role of corrections due to the nonideal effects.

As a matter of fact, just as in the numerical experiment where the OCP contributions are the leading terms, in the present model, quantitatively the Madelung energy is by far the most dominant part in the interaction energy [Eq. (1)], especially in the strong coupling regime ($\Gamma > 2$). Moreover, Madelung contribution to the internal energy as well as to the pressure have the same behavior as the OCP quantities, where some values, for a wide range of coupling strength are given in Table I. The relative deviation between the two models is seen to be small, less than 10 %, and particularly as $\Gamma$ grows ($\Gamma > 10$).

$U_M$ and $P_M$ are defined as:

$$\frac{U_M}{N\kappa_B T} = \Gamma \left[ \frac{\partial}{\partial \Gamma} \left( \frac{E_M}{N\kappa_B T} \right) \right]_{r_s}, \tag{6}$$

and

$$\frac{P_M}{n\kappa_B T} = \frac{1}{3} \frac{U_M}{N\kappa_B T} - \frac{r_s}{3} \left[ \frac{\partial}{\partial r_s} \left( \frac{E_M}{N\kappa_B T} \right) \right]_{r_s}, \tag{7}$$

To develop further the quantitative analysis, a comparison is presented showing the results of the present model and those of the most existing theories which have been analyzed and some of them improved in a comprehensive work by Galam and Hansen [7].

In Table II, a comparison of the thermodynamic functions is reported for a wide range of coupling strengths.

These functions $-\Delta P / n\kappa_B T$ and $-\Delta U / N\kappa_B T$ are defined by replacing in Eqs (6) and (7), $U_M$ and $P_M$ by $\Delta U$ and $\Delta P$ respectively and $E_M$ by $\Delta F$ defined in Eq. (1).

In Table II, some of the notations are borrowed from Ref. [7] for each value of $\Gamma$ are reported in the first line $-\Delta P / n\kappa_B T$ and the second line $-\Delta U / N\kappa_B T$ as calculated in the present model and compared to those obtained by Monte Carlo simulations of DeWitt and Hubbard [16], perturbation expansion (pert.) and also $q_{TF}$ expansion (where $q_{TF}$ is the dimensionless Thomas Fermi wave number), (Var. OCP) are results of a variational method based on Gibbs-Bogolyobov inequality and the physical idea of an effective charge reduction of the ions. The last two columns are the results of the hard sphere variational approach (Var. HS1) while (Var. HS2) are the results of the same method proposed by Ross and Seale [17].

The general statement which can be made on the value of the excess pressure and the internal energy reported in Table II is that they are globally very close to each other. The slight discrepancies between the different models are still contained within few percents.

Moreover, the present model presents the advantage by the analytical feature of the results to obtain the thermal parameters such as heat capacity adiabatic exponents which can be easily deduced as follows:

The heat capacity at constant volume is

$$\frac{C_v}{N\kappa_B} = \frac{3}{2} - \Gamma^2 \frac{\partial}{\partial \Gamma}\left[\left(\frac{\Delta U}{N\kappa_B T}\right)/\Gamma\right], \quad (8)$$

and the generalized Gruneissen as:

$$\gamma = \left(\frac{\partial \ln T}{\partial \ln n}\right)_s \quad (9)$$

The isothermal compressibility is defined as

$$\Delta K = \left(n \frac{\partial}{\partial n}(\Delta P)_T\right)^{-1}. \quad (10)$$

These parameters will be analyzed in a future work.

**Table I**. Excess pressure and internal energy due to Madelung contributions of the present model are shown to behave as the OCP quantities and are in good agreement especially as $\Gamma$ grows.

| $\Gamma$ | 2 | 6 | 10 | 20 | 40 | 70 | 100 | 120 | 140 | 160 |
|---|---|---|---|---|---|---|---|---|---|---|
| $\theta$ | 0.2715 | 0.0905 | 0.0543 | 0.0271 | 0.0136 | 0.0078 | 0.0054 | 0.0045 | 0.0039 | 0.0034 |
| **Present** $P_M/nk_BT$ | -0.60 | -1.80 | -3.00 | -6 | -12 | -21 | -30 | -36 | -42 | -48 |
| $U_M/Nk_BT$ | -1.80 | -5.40 | -9.00 | -18 | -36 | -63 | -90 | -108 | -126 | -144 |
| OCP | -0.444 | -1.53 | -2.66 | -5.55 | -11.41 | -20.27 | -29.27 | -35.09 | -41.03 | -46.97 |
|  | -1.332 | -4.59 | -7.99 | -16.66 | -34.24 | -60.81 | -87.47 | -105.3 | -123.1 | -140.9 |

Table II. For each Γ and θ are reported the excess pressure $-\Delta P/nk_BT$ (first line) and the excess internal energy $-\Delta U/Nk_BT$ (second line) as calculated in the present model along with those obtained by other methods for $r_s=0.1$.

| Γ | θ $r_s = 0.1$ | Present theory | Monte Carlo | $q_{TF}$ expansion | pert. | Var. OCP | Var. HSI. | Var.HS2 |
|---|---|---|---|---|---|---|---|---|
| 2 | 0.02715 | 0.280 | 0.442 | 0.437 | 0.438 | 0.436 | 0.338 | 0.370 |
|   |         | 1.320 | 1.358 | 1.363 | 1.370 | 1.360 | 1.139 | 1.442 |
| 6 | 0.00905 | 1.490 | 1.524 | 1.516 | 1.516 | 1.517 | 1.386 | 1.425 |
|   |         | 4.936 | 4.645 | 4.647 | 4.647 | 4.645 | 4.367 | 4.463 |
| 10 | 0.00543 | 2.697 | 2.573 | 2.565 | 2.565 | 2.567 | 2.498 | 2.534 |
|    |         | 8.546 | 8.071 | 8.069 | 8.072 | 8.068 | 7.753 | 7.854 |
| 20 | 0.00271 | 5.709 | 5.540 | 5.530 | 5.530 | 5.532 | 5.353 | 5.384 |
|    |         | 17.56 | 16.80 | 16.79 | 16.80 | 16.79 | 16.41 | 16.51 |
| 40 | 0.00136 | 11.73 | 11.38 | 11.38 | 11.38 | 11.38 | 11.17 | 11.29 |
|    |         | 35.59 | 34.50 | 34.271 | 34.48 | 34.41 | 34.00 | 34.10 |
| 70 | 0.00077 | 20.74 | 20.24 | 20.23 | 20.23 | 20.23 | 19.98 | 20.02 |
|    |         | 62.61 | 61.24 | 61.18 | 61.22 | 61.18 | 60.62 | 60.72 |
| 100 | 0.00054 | 29.76 | 29.12 | 29.11 | 29.11 | 29.11 | 28.84 | 28.87 |
|     |         | 89.63 | 88.07 | 87.98 | 88.05 | 87.97 | 87.36 | 87.46 |
| 120 | 0.00045 | 35.76 | 35.06 | 35.04 | 35.05 | 35.00 | 34.75 | 34.79 |
|     |         | 107.6 | 106.0 | 105.8 | 106.0 | 105.8 | 105.2 | 105.3 |
| 140 | 0.00039 | 41.77 | 41.00 | 40.94 | 40.98 | 40.98 | 40.68 | 40.72 |
|     |         | 125.6 | 123.9 | 123.8 | 123.9 | 123.8 | 123.1 | 12.32 |
| 160 | 0.00034 | 47.78 | 46.94 | 46.92 | 46.95 | 46.97 | 46.68 | 45.65 |
|     |         | 143.7 | 141.8 | 141.6 | 142.0 | 141.7 | 141.01 | 141.1 |